\documentclass[11pt]{revtex4}
\usepackage[utf8]{inputenc}
\usepackage[T1]{fontenc}
\usepackage[english]{babel}
\usepackage{amssymb,amsfonts,amsmath}

\usepackage{graphicx}
\usepackage{color}
\usepackage{bm}
\def\newpic#1{}

\begin{document}
\title{Theory of elastic interaction between axially symmetric 3D skyrmions in confined chiral nematic liquid crystals and in skyrmion bags}

\author{S.~B.~Chernyshuk $^{1}$ and Ye.~G.~Rudnikov $^{2,3}$}

\affiliation{ $^{1}$ Institute of Physics, NAS Ukraine, Prospekt Nauki 46, Kyiv 03028, Ukraine }
\affiliation{ $^{2}$ Department of Physics, Taras Shevchenko National University of Kyiv, Ukraine; 2 Glushkov Prosp., Build. 1, Kyiv, 03680, Ukraine}
\affiliation{ $^{3}$ National Technical University of Ukraine ``Igor Sikorsky Kyiv Polytechnic Institute'', 37, Prosp.Peremohy, Solomyanskyi district, Kyiv, 03056, Ukraine }

\thanks{emails: stasubf@gmail.com, rudnikof@yahoo.com}

\begin{abstract}
We study axially symmetrical 3D skyrmionic particles (torons, hopfions) in a thin homeotropic cell filled with a cholesteric liquid crystal. In this case, the contribution of the chirality term on far distances becomes small and a 3D skyrmion can be described as a very specific particle in nematic with multipole moments of different chiralities. We show that small skyrmions asymptotically can be presented as a particle with six multipole moments. Thus, we find the exact analytical solution for the far field director configuration around the 3D skyrmion and the exact analytical expression for the elastic interaction potential between 3D skyrmions in a homeotropic cell with thickness $L$. The comparison with the experiment allows us to recover approximately multipole coefficients of one small 3D skyrmion. We also describe the expansion of the skyrmion bag analytically depending on the number $N$ of anti-skyrmions in it and obtain an agreement with the experiment.

Along the way, we determined the director field around a chiral axially symmetric colloidal particle in the unlimited nematic liquid crystal and in the confined homeotropic cell, taking into account higher order elastic terms with orders $l=1\div 6$. We also have found the elastic interaction potential between such particles.
\end{abstract}
\maketitle

\section{Introduction}
Skyrmions are topological solitons, built from configurations of various fields with chirality.

Originally, Skyrme built an effective Lagrangian entirely in meson fields (pions), and baryons -- as fermions -- could emerge from that Lagrangian as solitons \cite{sk1,sk2}. Classical solitons of minimal energy for each baryon number were called Skyrmions. They describe fairly well not only the `elementary' nucleons (proton and neytron) but the finite nuclei with a mass number up to 22 as well, with predictions of certain ground state properties that have not been revealed by the standard many-body approaches developed in nuclear theory \cite{sk3,sk4}.
Nowadays, the concept of skyrmions as multidimensional topological solitons plays an important role in many branches of modern physics,  in particular in condensed matter systems \cite{sk3}. Usually such systems are characterized by a lack of inversion symmetry (that is called chirality).

Skyrmions were predicted and observed in quantum Hall systems \cite{mag1}-\cite{mag3}. Later, Bose--Einstein condensates with spin degrees of freedom were shown to accommodate Skyrmions \cite{mag4}-\cite{mag6}. As well, great attention has been paid to Skyrmions in noncentrosymmetric ferromagnets \cite{mag7}-\cite{mag13} with the presence of Dzyalonshinskii-Moriya spin-orbit interaction, such as MnSi \cite{mag3},\cite{mag8,mag9} and $Fe_{1-x}Co_xSi$ \cite{mag10,mag11}. Chiral magnetic skyrmions are now considered as promising objects for applications in magnetic data storage technologies and in the emerging spin transport electronics (spintronics), because local twisted magnetic structures coupled to electric or spin currents could be used to manipulate electrons and their spins \cite{mag12},\cite{mag13}. Theoretical aspects of skyrmions in chiral magnets were investigated in \cite{magtheory1}-\cite{magtheory10}.

Recently, LC skyrmions have been realized as micron sized solitons in a chiral nematic material confined between two parallel substrates and they attracted great research interest \cite{lcskyr1}-\cite{lcskyr27}.

It is not a secret that 3D skyrmions in liquid crystals are similar to ordinary particles. On the other hand, the behavior of colloidal particles in liquid crystals is a very interesting topic \cite{lccolloid1}-\cite{lccolloid6}. Novel anisotropic elastic interactions between colloidal particles lead to the formation of various structures and even colloidal 2D and 3D quasicrystals \cite{lccolloidstructures1}-\cite{lccolloidstructures13}. The theoretical aspects of the behavior of colloidal particles in nematic liquid crystals are studied in \cite{lccolloidtheory1}-\cite{lccolloidtheory14}, and the theoretical description of 2D and 3D colloidal quasicrystals is given in \cite{lccolloidtheory15}-\cite{lccolloidtheory17}.

In this paper, we will try to bridge the gap between 3D skyrmions and colloidal particles in liquid crystals. We will look at the skyrmion as a colloidal particle and consider the interaction between skyrmions as an elastic interaction between colloidal particles. This can be done exactly outside the core of the skyrmion, beyond the region with strong director deformations. Then the director distribution can be accurately determined analytically and the elastic interaction potential between 3D skyrmion particles can be calculated exactly. We consider skyrmions with axial symmetry in a homeotropic cell of thickness $L$ with strong boundary conditions, when the director looks vertically at the surfaces of the bounding planes. It turns out that the chirality of the liquid crystal manifests itself in the birth of chiral multipole coefficients $a_l^{\gamma}$ , which are absent in ordinary colloidal particles in the nematic. Usually, colloidal particles in nematics have only zero-helicity multipole coefficients $a_l^{0}$. And then it turns out that the 3D skyrmion in the region of weak deformations can be described as a set of multipoles: zero $a_l^{0}$ and chiral $a_l^{\gamma}$. We consider in more detail the case of 3D skyrmion particles that are symmetric in shape with respect to the median horizontal plane $z=L/2$ and consider only the helicity $\gamma=\pi/2$. We also restrict ourselves to multipole moments up to the 6-th order inclusively. Then we obtain the interaction potential between 3D skyrmions and find the multipole coefficients from agreement with experiment.

Along the way, we determine the director field in the unlimited nematic around a chiral axially symmetric colloidal particle, taking into account higher order elastic terms, and also find the elastic interaction potential between such particles with multipole moments of orders  $l=1\div 6$.

Next, we apply our results to 3D skyrmions in the skyrmion bag. As the number $N$ of skyrmions in the bag increases, the distance between them gets larger. We find an approximate analytical formula for the distance between skyrmions as a function of the number $N$ of skyrmions in the bag and cell thickness $L$, which also agrees with the experiment.

\section{3D skyrmions as a colloidal particle: a new perspective}

In this work, we will not take into account the influence of an electric or magnetic field.
Free energy of the system in the one-constant approximation is equal to:

\begin{equation}\label{Frank_FE}
F_{bulk} = \frac{K}{2} \int dV \left[ (div \textbf{n} )^2 +(\textbf{n} rot\textbf{n}+q_0)^2+ (\textbf{n} \times rot\textbf{n} )^2 \right],
\end{equation}

where $K$ is the elastic constant and $q_0=2\pi/p$ is the wavenumber of the cholesteric helix in the ground state, and $p$ is the pitch of the helix at which the director completes a revolution of $2\pi$ .The chiral term $Kq_0\textbf{n} rot\textbf{n}$ arises due to the presence of chiral molecules in which there is no inversion center. We consider the geometry when a cholesteric liquid crystal or a nemato-cholesteric mixture is placed into a homeotropic cell with a thickness $L$ of the order of the helix pitch $L/p\approx1$. We will assume that rigid boundary conditions $\textbf{n} ||\textbf{z}$ are fixed on the cell boundaries.

\begin{figure}
\begin{center}
\includegraphics[width=0.7\columnwidth]{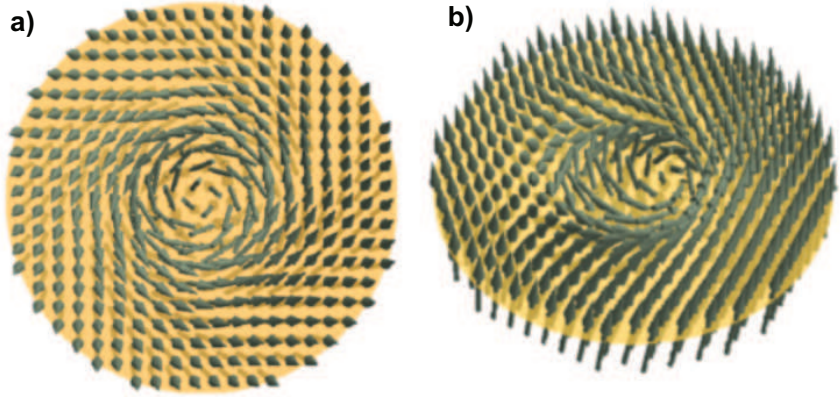}
\caption{Schematic representation of a 2D skyrmion. The director makes a full turn by $\pi$ when moving from the center to the periphery. Here a) top view and b) side view.}\label{skyrmion2d}
\end{center}
\end{figure}

In such systems, there can arise axially symmetric 3D solitons (torons, hopfions), which for generalization we will call 3D skyrmions (or simply skyrmions) in a liquid crystal, although usually in the condensed matter physics the word "skyrmion" refers to a 2D soliton shown in the figure ( \ref{skyrmion2d}). Due to the homeotropic boundary conditions, the director far from the skyrmion tends to the equilibrium distribution of $\textbf{n}_{0} ||\textbf{z}$ in the entire space. Regardless of the shape, size, and other parameters of the 3D skyrmion, the director field deformations decay with distance from the skyrmion and eventually become small, $\textbf{n}(\textbf{r})$ approximately $(n_{x}, n_ {y} , 1)$, where $|n_{\mu}| \ll 1$, $\textbf{n}_{0} = (0, 0, 1)$ is the ground state, $\mu = \{ x,y \}$ hereinafter.
We will consider just this zone, where the deformation is already small. Assuming $|n_{\mu}|\sim \epsilon$ one can easily obtain that up to $\epsilon^2$ chiral term $\textbf{n} rot\textbf{n}=\partial_z(n_x n_y)-2n_x \partial_z n_y+\partial_x n_y-\partial_y n_x$. In this paper, we consider exclusively axially symmetric skyrmions. Obviously, for axially symmetric skyrmions, the space volume integral of this distribution equals zero $\int\textbf{n} rot\textbf{n}dV=0$, i.e. far from such a skyrmion, a cholesteric liquid crystal in a homeotropic cell can be considered as a nematic. That is a very important point!

Thus, far from the skyrmion the bulk free energy takes a simple harmonic form
\begin{equation}\label{F_bulk}
F_{\text{bulk}} = \frac{K}{2} \int dV \nabla n_{\mu} \cdot \nabla n_{\mu},
\end{equation}
where repeated $\mu$ implies summation over $x$ and $y$, \textit{i.e.} $(\nabla n_{\mu})^{2}=(\nabla n_{x})^{2} + (\nabla n_{y})^{2}$.
The Euler-Lagrange equations for $n_{\mu}$ are of the Laplace type
\begin{equation}\label{Laplace_eq}
\Delta n_{\mu} = 0.
\end{equation}

\subsection{Colloidal particles with chirality in the unlimited nematic liquid crystal}

Now, for clarity, we will temporarily digress from the LC cell under consideration and see what axially symmetric solutions can arise in an unlimited space. That is, suppose that there is some small axially symmetric particle in the unlimited nematic liquid crystal.

In the paper \cite{lccolloidtheory13} it was shown that the solution of the Laplace equation for axially symmetric particles has the form:
\begin{equation}
n_{\mu}=\sum^{N}_{l=1}a_{l}(-1)^{l}\partial_{\mu}\partial_{z}^{l-1}\frac{1}{r}  \label{hn}
\end{equation}
where $a_{l}$ is the multipole moment of the order $l$ and $2^{l}$ is the multipolarity; $N$ - is the maximum possible order without anharmonic corrections. In this case $n_x\propto cos(\varphi)$ and $n_y\propto sin(\varphi)$ with $\varphi$ being the polar angle. But this result is valid only in the absence of twisting around the particle. For example
particles with a hyperbolic hedgehog, a Saturn-ring configuration, boojums, or hexadecapole particles in \cite{lccolloidtheory14} capture the formula (\ref{hn}).

In order to find the energy of the system: particle(s) + LC , it is necessary to introduce some effective free energy functional $F_{eff}$ \cite{lccolloidtheory13}, so that it's Euler-Lagrange equations would have the above solutions (\ref{hn}). In the one constant approximation with Frank constant $K$ the effective functional has the form:
\begin{equation}
F_{eff}=K\int d^{3}x\left\{\frac{(\nabla n_{\mu})^{2}}{2}-4\pi\sum^{N}_{l=1}A_{l}(\textbf{x})\partial_{\mu}\partial_{z}^{l-1}n_{\mu} \right\}\label{flin}
\end{equation}
which brings Euler-Lagrange equations:
\begin{equation}
\Delta n_{\mu}=4\pi\sum^{N}_{l=1}(-1)^{l-1}\partial_{\mu}\partial_{z}^{l-1}A_{l}(\textbf{x})\label{nmu}
\end{equation}
where $A_{l}(\textbf{x})$ are multipole moment densities, $\mu=x,y$ and repeated $\mu$ means summation on $x$ and $y$ like $\partial_{\mu}n_{\mu}=\partial_{x}n_{x}+\partial_{y}n_{y}$.
For the infinite space the solution has the known form:
\begin{equation}
n_{\mu}(\textbf{x})=\int d^{3}\textbf{x}' \frac{1}{\left|\textbf{x}-\textbf{x}'\right|}\sum^{N}_{l=1}(-1)^{l}\partial_{\mu}'\partial_{z}'^{l-1}A_{l}(\textbf{x}') \label{solmain}
\end{equation}

If we consider $A_{l}(\textbf{x})=a_{l}\delta(\textbf{x})$ this really brings solution (\ref{hn}). This means that effective functional (\ref{flin}) correctly describes the interaction between the particle and LC.

 Consider $N_{p}$ particles in the NLC, so that $A_{l}(\textbf{x})=\sum_{i}a_{l}^{i}\delta(\textbf{x}-\textbf{x}_{i})$, $i=1\div N_{p}$ . Then substitution (\ref{solmain}) into $F_{eff}$ (\ref{flin}) brings: $F_{eff}=U^{self}+U^{interaction}$ where $U^{self}=\sum_{i}U_{i}^{self} $ , here $U_{i}^{self}$ is the divergent self energy.\\
Interaction energy $ U^{interaction}=\sum_{i<j}U_{ij}^{int} $.  Here $U_{ij}^{int}$ is the elastic interaction energy between $i$ and $j$ particles in the unlimited NLC:
\begin{equation}
U_{ij}^{int}=4\pi K \sum^{N}_{l,l'=1}a_{l}a_{l'}'(-1)^{l'}\frac{(l+l')!}{r^{l+l'+1}}P_{l+l'}(\cos\theta)\label{uint}
\end{equation}
Here unprimed quantities $a_{l}$ are used for particle $i$ and primed $a_{l'}'$ for particle $j$, $r=|\textbf{x}_{i}-\textbf{x}_{j}|$, $\theta$ is the angle between $\textbf{r}$ and z and we used the relation $P_{l}(\cos\theta)=(-1)^{l}\frac{r^{l+1}}{l!}\partial_{z}^{l}\frac{1}{r}$ for Legendre polynomials $P_{l}$. This is the general expression for the elastic interaction potential between axially symmetric colloidal particles in the unlimited NLC when taking into account the high order elastic terms \cite{lccolloidtheory13}.

But what will happen if there is still helical twisting or circular cutting around the particle? In such a more general case, the axisymmetric solution has the property $n_x^\gamma\propto cos(\varphi+\gamma)$ and $n_y^\gamma\propto sin(\varphi+\gamma)$ with $\varphi$ being the polar angle and $\gamma$ being the helicity number. Then a more general axially symmetric solution with helicity $\gamma$ can be represented as a linear combination of solutions (\ref{hn}) in the form:

\begin{equation}
n_{x}^\gamma=\sum^{N}_{l=1}a_{l}^\gamma(-1)^{l}(cos\gamma\partial_{x}\partial_{z}^{l-1}-sin\gamma\partial_{y}\partial_{z}^{l-1})\frac{1}{r}\propto cos(\varphi+\gamma)  \label{hnxg}
\end{equation}
\begin{equation}
n_{y}^\gamma=\sum^{N}_{l=1}a_{l}^\gamma(-1)^{l}(sin\gamma\partial_{x}\partial_{z}^{l-1}+cos\gamma\partial_{y}\partial_{z}^{l-1})\frac{1}{r}\propto sin(\varphi+\gamma)  \label{hnyg}
\end{equation}
and the most general axially symmetric solution is the sum over all possible helicities$\gamma$:
\begin{equation}
n_{x}=\sum_\gamma n_{x}^\gamma=\sum^{N}_{l=1}(-1)^{l}(\sum_\gamma a_{l}^\gamma cos\gamma\partial_{x}\partial_{z}^{l-1}-\sum_\gamma a_{l}^\gamma sin\gamma\partial_{y}\partial_{z}^{l-1})\frac{1}{r} \label{hnx}
\end{equation}
\begin{equation}
n_{y}=\sum_\gamma n_{y}^\gamma=\sum^{N}_{l=1}(-1)^{l}(\sum_\gamma a_{l}^\gamma sin\gamma\partial_{x}\partial_{z}^{l-1}+\sum_\gamma a_{l}^\gamma cos\gamma\partial_{y}\partial_{z}^{l-1})\frac{1}{r} \label{hny}
\end{equation}

If we introduce such a pair of orthogonal vectors $\vec{A}_l^x$=$(\sum_\gamma a_{l}^\gamma cos\gamma,-\sum_\gamma a_{l }^\gamma sin\gamma)=(a_l^x,a_l^y)$ and $\vec{A}_l^y$=$(\sum_\gamma a_{l}^\gamma sin\gamma,\sum_ \gamma a_{l}^\gamma cos\gamma)=(-a_l^y,a_l^x)$, and $\vec{A}_l^x\cdot\vec{A}_l^y=0$ for each multipole order $l$, then the formulas will be rewritten in the following form:

\begin{equation}
n_{\mu}=\sum^{N}_{l=1}(-1)^{l}A_l^{\mu\alpha}\partial_{\alpha}\partial_{z}^{l-1}\frac{1}{r} \label{hnmu_chiral}
\end{equation}

If we introduce such multipole moment densities $A_l^{\mu\alpha}(\textbf{x})=A_l^{\mu\alpha}\delta(\textbf{x})$, then the free energy functional (\ref{flin}) will be rewritten as:
\begin{equation}
F_{eff}=K\int d^{3}x\left\{\frac{(\nabla n_{\mu})^{2}}{2}-4\pi\sum^{N}_{ l=1}A_l^{\mu\alpha}(\textbf{x})\partial_{\alpha}\partial_{z}^{l-1}n_{\mu} \right\}\label{flinhelicity}
\end{equation}

This functional correctly describes the interaction between a liquid crystal and an axially symmetric particle with chirality since it's Euler-Lagrange equation is:

\begin{equation}
\Delta n_{\mu}=4\pi\sum^{N}_{l=1}(-1)^{l-1}\partial_{\alpha}\partial_{z}^{l-1}A_{l}^{\mu\alpha}(\textbf{x})\label{nmuhelical}
\end{equation}

and for the infinite space it has the known solution :
\begin{equation}
n_{\mu}(\textbf{x})=\int d^{3}\textbf{x}' \frac{1}{\left|\textbf{x}-\textbf{x}'\right|}\sum^{N}_{l=1}(-1)^{l}\partial_{\alpha}'\partial_{z}'^{l-1}A_{l}^{\mu\alpha}(\textbf{x}') \label{solmainhelcal}
\end{equation}
which coinsides with (\ref{hnmu_chiral}).

If we consider $N_{p}$ particles with helicity in the NLC, so that $A_{l}^{\mu\alpha}(\textbf{x})=\sum_{i}A_{l,i}^{\mu\alpha}\delta(\textbf{x}-\textbf{x}_{i})$, $i=1\div N_{p}$ . Then substitution (\ref{nmuhelical}) into $F_{eff}$ (\ref{flinhelicity}) brings: $F_{eff}=U^{self}+U^{interaction}$ where $U^{self}=\sum_{i}U_{i}^{self} $ , here $U_{i}^{self}$ is the divergent self energy.\\
Interaction energy $ U^{interaction}=\sum_{i<j}U_{ij}^{int} $.  Here $U_{ij}^{int}$ is the elastic interaction energy between $i$ and $j$ particles in the unlimited NLC:
\begin{equation}
U_{ij}^{int}=4\pi K \sum^{N}_{l,l'=1}(a_{l}^x a_{l'}'^x+a_{l}^y a_{l'}'^y)(-1)^{l'}\frac{(l+l')!}{r^{l+l'+1}}P_{l+l'}(\cos\theta)\label{uinthel}
\end{equation}
Here unprimed quantities $(a_{l}^x,a_{l}^y)$ are used for particle $i$ and primed $(a_{l}'^x,a_{l}'^y)$ for particle $j$, $r=|\textbf{x}_{i}-\textbf{x}_{j}|$, $\theta$ is the angle between $\textbf{r}$ and $z$ . In general case :
\begin{equation}
a_{l}^x a_{l'}'^x+a_{l}^y a_{l'}'^y=\sum_{\gamma\gamma'}a_{l}^{\gamma} a_{l'}^{\gamma'}\,'cos(\gamma-\gamma')\label{cosgamma}
\end{equation}
with summation over all possible helicities $\gamma$ for every multipole order $l$. Actually in general case helical twisting of the molecules on the particle' surface can produce different helicities and different multipole coefficients $a_{l}^{\gamma}$. But we do not need to know all $a_{l}^{\gamma}$, as it is sufficient to know only two coefficients $(a_l^x,a_l^y)$ for each $l$ to determine the director field: $(a_l^x,a_l^y)=(\sum_\gamma a_{l}^\gamma cos\gamma,-\sum_\gamma a_{l}^\gamma sin\gamma)$. When helicity or spiral cutting on the surface is absent, only $a_{l}^{0}\equiv a_{l}$ remain.

Expression (\ref{uinthel}) is the general expression for the elastic interaction potential between axially symmetric helical colloidal particles in the unlimited NLC with taking into account the high order elastic terms.

\subsection{3D skyrmion as a colloidal particle with chirality in confined nematic liquid crystal}

 In the case of confined NLC it means the replacement of $\frac{1}{r}=\frac{1}{|\textbf{x}_{i}-\textbf{x}_{j}|}$ with Green's function $G(\textbf{x},\textbf{x}')$ (see \cite{lccolloidtheory8,lccolloidtheory9}), which satisfies equation $\Delta_{\textbf{x}}G(\textbf{x},\textbf{x}')=-4\pi \delta(\textbf{x}-\textbf{x}')$ for  $\textbf{x},\textbf{x}'\in \textbf{V}$ ($\textbf{V}$ is the volume of the bulk NLC) and $G(\textbf{x},\textbf{s})=0 $ for any $\textbf{s}$ of the bounding surfaces $\Sigma$.
Then director field in any point $\textbf{x}$ inside the cell is defined as:
\begin{equation}
n_{\mu}(\textbf{x})=\int d^{3}\textbf{x}' G(\textbf{x},\textbf{x}')\sum^{N}_{l=1}(-1)^{l}\partial_{\alpha}'\partial_{z}'^{l-1}A_{l}^{\mu\alpha}(\textbf{x}') \label{solmainhelcalcell}
\end{equation}

Then formula (\ref{uint}) for the confined NLC has the form:
\begin{equation}
U_{ij}^{int,confined}=-4\pi K\sum^{N}_{l,l'=1}(a_{l}^x a_{l'}'^x+a_{l}^y a_{l'}'^y)\partial_{\mu}\partial_{\mu}'\partial_{z}^{l-1}\partial_{z}'^{l'-1}G(\textbf{x}_{i},\textbf{x}_{j}')\label{uintconfhelical}
\end{equation}

The first most important dipole term in the general case has the form:

\begin{equation}
U_{ij,dipole}^{int,confined}=-4\pi K(a_{1}^x a_{1'}'^x+a_{1}^y a_{1'}'^y)\partial_{\mu}\partial_{\mu}'G(\textbf{x}_{i},\textbf{x}_{j}')=-4\pi K\sum_{\gamma\gamma'}a_{1}^{\gamma} a_{1'}^{\gamma'}\,'cos(\gamma-\gamma')\partial_{\mu}\partial_{\mu}'G(\textbf{x}_{i},\textbf{x}_{j}')\label{uintconfhelicaldipole}
\end{equation}

Formula (\ref{uintconfhelical}) gives the most general expression for the elastic interaction energy between any axially symmetric colloidal particles  with helicity or between 3D skyrmions in the homeotropic cell with the thickness $L$.

Now we can return to the consideration of one skyrmion (toron, hopfion) inside a homeotropic cell with thickness $L$. In the general case, it is described by $2N$ numbers $(a_l^x,a_l^y)$ - multipole moments, $l=1\div N$. But symmetry considerations make it possible to reduce the number of these coefficients. First, below we will restrict ourselves to cases with helicity $\gamma=0$ and $\gamma=\pi/2$ (for example, we will not consider hopfions with $\gamma=\pi/4$). Then all coefficients $a_l^0$ and $a_l^{\pi/2}$ can be non-zero. For instance, for any axially symmetric 3D solitons with an asymmetric shape with respect to the median plane $z=L/2$ or similar particles (for example, a dipole particle with a hyperbolic hedgehog in a homeotropic cell with a cholesteric at $L/p\approx1$), all the coefficients $a_l^0$ and $a_l^{\pi/2 }$ are different from zero.

Secondly, it can be shown (we will not give a proof of this here) that if a 3D soliton has a symmetric shape with respect to the middle of the cell $z=L/2$, then only even coefficients $a_l^0$ and odd coefficients $a_l^ {\pi/2}$ remain. Thus, for $z$-symmetric 3D solitons, only $N$ non-zero numbers remain. In this paper, we will consider only terms up to $N=6$ and will consider only such 3D skyrmions that have symmetrical shape with respect to the median plane $z=L/2$. Then only odd helical coefficients $(a_1^{\pi/2},a_3^{\pi/2},a_5^{\pi/2} )$ - chiral dipole, chiral octupole and chiral moment of the order 5 will be nonzero. And also there will be even nonzero coefficients $a_2^0,a_4^0,a_6^0$ - ordinary quadrupole moment, hexadecapole moment and sixth order moment. For example, such moments are present in particles with boojums \cite{lccolloidtheory13} and in hexadecapole colloids \cite{lccolloidtheory14}.

Let's consider a 3D skyrmion, which has a symmetrical shape with respect to the median plane $z=L/2$, and which is located at the center of a homeotropic cell at the point $x_0=(0,0,L/2)$. Then the director field (\ref{solmainhelcalcell}) at large distances from the skyrmion center is described by the formula

\begin{equation}
n_{\mu}(\textbf{x})=\sum^{N=6}_{l=1}A_{l}^{\mu\alpha} \partial_{\alpha}'\partial_{z}'^{l-1}G(\textbf{x},\textbf{x}')|_{x'\rightarrow (0,0,L/2)} \label{solmainhelcalcell1}
\end{equation}

The Green function for the homeotropic cell can be taken from electrodynamics \cite{jac}:

\begin{equation}
G(\textbf{x},\textbf{x}')= \frac{4}{L}\sum_{n=1}^{\infty} sin(\frac{n\pi z}{L})sin(\frac{n\pi z'}{L})K_{0}(\frac{ n\pi\rho}{L})\label{greenf}
\end{equation}
where $\rho=\sqrt{(y-y^{\prime})^{2}+(x-x^{\prime})^{2}}$ is the horizontal projection of the distance between particles and $z\in[0,L]$.

Then the director field at an arbitrary point $\textbf{x}$ with coordinates $(\rho, \varphi,z)$ inside the cell can be written as:

\begin{equation}\label{solmainhelcalcellx1}
\begin{gathered}
n_{x}(\textbf{x})=\frac{4}{L}cos(\varphi+\pi/2)\sum_{n=1,3,5..}^{\infty} sin(n\pi/2) sin(\frac{n\pi z}{L}) K_{1}(\frac{ n\pi\rho}{L})\left[a_1^{\pi/2}\left(\frac{n\pi}{L}\right)-a_3^{\pi/2}\left(\frac{n\pi}{L}\right)^3+a_5^{\pi/2}\left(\frac{n\pi}{L}\right)^5\right] +\\
\\ +\frac{4}{L}cos(\varphi)\sum_{n=2,4,6..}^{\infty} cos(n\pi/2) sin(\frac{n\pi z}{L}) K_{1}(\frac{ n\pi\rho}{L})\left[a_2^{0}\left(\frac{n\pi}{L}\right)^2-a_4^{0}\left(\frac{n\pi}{L}\right)^4+a_6^{0}\left(\frac{n\pi}{L}\right)^6\right]
\end{gathered}
\end{equation}
\begin{equation}\label{solmainhelcalcelly1}
\begin{gathered}
n_{y}(\textbf{x})=\frac{4}{L}sin(\varphi+\pi/2)\sum_{n=1,3,5..}^{\infty} sin(n\pi/2) sin(\frac{n\pi z}{L}) K_{1}(\frac{ n\pi\rho}{L})\left[a_1^{\pi/2}\left(\frac{n\pi}{L}\right)-a_3^{\pi/2}\left(\frac{n\pi}{L}\right)^3+a_5^{\pi/2}\left(\frac{n\pi}{L}\right)^5\right] +\\
\\ +\frac{4}{L}sin(\varphi)\sum_{n=2,4,6..}^{\infty} cos(n\pi/2) sin(\frac{n\pi z}{L}) K_{1}(\frac{ n\pi\rho}{L})\left[a_2^{0}\left(\frac{n\pi}{L}\right)^2-a_4^{0}\left(\frac{n\pi}{L}\right)^4+a_6^{0}\left(\frac{n\pi}{L}\right)^6\right]
\end{gathered}
\end{equation}

Let $d$ be the diameter of a skyrmion in the symmetry plane $z=L/2$ and we will introduce the dimensionless $\sigma=d/L$. If we introduce dimensionless multipole moments ($b_1,b_2,b_3,b_4,b_5,b_6$) , then the coefficients can be represented as $(a_1^{\pi/2},a_3^{\pi/2},a_5^ {\pi/2} )=(b_1d^2,b_3d^4,b_5d^6 )$ and $(a_2^{0},a_4^{0},a_6^{0} )=(b_2d^3,b_4d ^5,b_6d^7 )$. Then, in the Cartesian coordinate system, the director field at any point $\textbf{x}$ with coordinates $(x, y,z)$ can be represented as:

\begin{equation}\label{nx2_sk}
\begin{gathered}
n_{x}(\textbf{x})=-\frac{4y}{\rho}\sum_{n=1,3,5..}^{\infty} sin(n\pi/2) sin(\frac{n\pi z}{L}) K_{1}(\frac{ n\pi\rho}{L})\left[b_1\left(n\pi\sigma\right)-b_3\left(n\pi\sigma\right)^3+b_5\left(n\pi\sigma\right)^5\right] +\\
\\ +\frac{4x}{\rho}\sum_{n=2,4,6..}^{\infty} cos(n\pi/2) sin(\frac{n\pi z}{L}) K_{1}(\frac{ n\pi\rho}{L})\left[b_2\left(n\pi\sigma\right)^2-b_4\left(n\pi\sigma\right)^4+b_6\left(n\pi\sigma\right)^6\right]
\end{gathered}
\end{equation}
\begin{equation}\label{ny2_sk}
\begin{gathered}
n_{y}(\textbf{x})=\frac{4x}{\rho}\sum_{n=1,3,5..}^{\infty} sin(n\pi/2) sin(\frac{n\pi z}{L}) K_{1}(\frac{ n\pi\rho}{L})\left[b_1\left(n\pi\sigma\right)-b_3\left(n\pi\sigma\right)^3+b_5\left(n\pi\sigma\right)^5\right] +\\
\\ +\frac{4y}{\rho}\sum_{n=2,4,6..}^{\infty} cos(n\pi/2) sin(\frac{n\pi z}{L}) K_{1}(\frac{ n\pi\rho}{L})\left[b_2\left(n\pi\sigma\right)^2-b_4\left(n\pi\sigma\right)^4+b_6\left(n\pi\sigma\right)^6\right]
\end{gathered}
\end{equation}

where $\rho=\sqrt{(y-y^{\prime})^{2}+(x-x^{\prime})^{2}}$. The formulas (\ref{nx2_sk}) and (\ref{ny2_sk}) define an analytical description of the director field for any axially symmetric skyrmion (3D soliton) outside it's strong deformation region, provided that it has a symmetric shape with respect to $z=L/2$ and it has no chirality other than $\gamma=\pi/2$. That is, different 3D skyrmionic particles (torons, hopfions) will differ only in sets of coefficients ($b_1,b_2,b_3,b_4,b_5,b_6$).

Consider now the interaction between two identical 3D skyrmionic particles. Since they are in the center, then $z=z'=L/2$ and let $\rho$ be the distance between them. Then the general formula (\ref{uintconfhelical}) gives the following interaction potential:

\begin{equation}
U_{ij}^{int,confined}=U_{ij}^{helical}+U_{ij}^{shape}\label{uintconfhelical1}
\end{equation}

\begin{equation}\label{uinthelical}
\begin{gathered}
U_{ij}^{helical}=16\pi \sigma Kd\sum_{n=1,3,5..}^{\infty}  K_{0}(\frac{ n\pi\rho}{L})\cdot[b_1^2\left(n\pi\sigma\right)^2-2b_1 b_3\left(n\pi\sigma\right)^4+2b_1 b_5\left(n\pi\sigma\right)^6 +\\
\\+ b_3^2\left(n\pi\sigma\right)^6-2b_3 b_5\left(n\pi\sigma\right)^8+b_5^2\left(n\pi\sigma\right)^{10} ]
\end{gathered}
\end{equation}
\begin{equation}\label{uinshape}
\begin{gathered}
U_{ij}^{shape}=16\pi \sigma Kd\sum_{n=2,4,6..}^{\infty} K_{0}(\frac{ n\pi\rho}{L})\cdot[b_2^2\left(n\pi\sigma\right)^4-2b_2 b_4\left(n\pi\sigma\right)^6+2b_2 b_6\left(n\pi\sigma\right)^8 +\\
\\+ b_4^2\left(n\pi\sigma\right)^8-2b_4 b_6\left(n\pi\sigma\right)^{10}+b_6^2\left(n\pi\sigma\right)^{12} ]
\end{gathered}
\end{equation}

\begin{figure}
\begin{center}
\includegraphics[width=0.6\columnwidth]{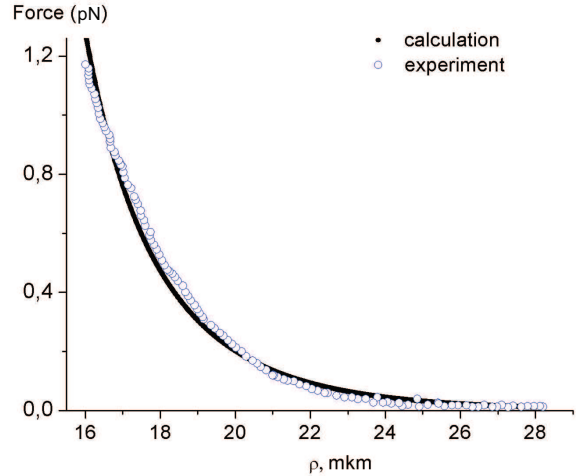}
\caption{The interaction force between 3D skyrmions in $pN$.Comparison between analytical expression (\ref{force}) with ($b_1,b_2,b_3,b_4,b_5,b_6$)=($0.17,-0.06,-0.003,0.0005, 0.0013,-0.00052$) and experiment. Experimental points were taken from \cite{lcskyr9}. }\label{pic_force}
\end{center}
\end{figure}

\begin{figure}
\begin{center}
\includegraphics[width=1.0\columnwidth]{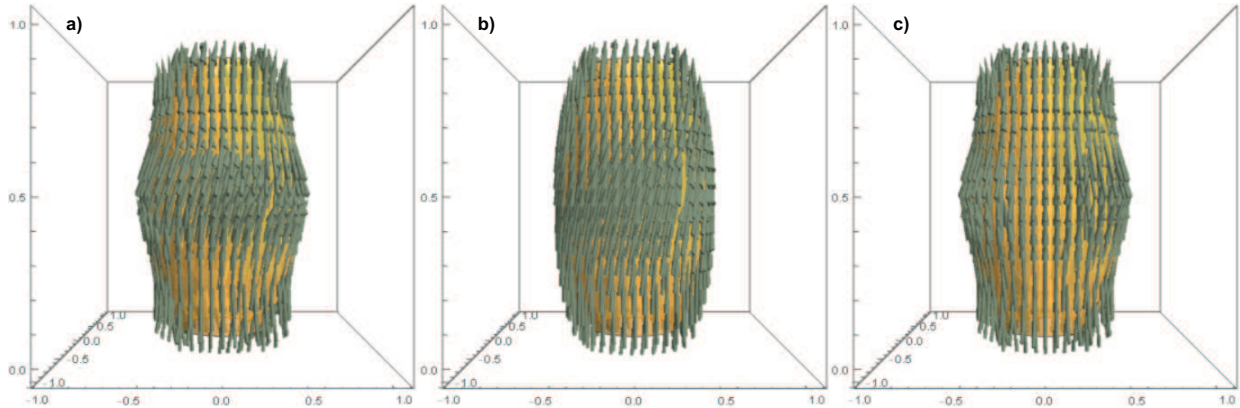}
\caption{Director field according to formulas (\ref{nx2_sk},\ref{ny2_sk}) on the spheroidal surface with radius 0.6L with first three terms in the series.  (a) Calculation with coefficients ($b_1,b_2,b_3,b_4,b_5,b_6$)=($0.17,-0.06,-0.003,0.0005, 0.0013,-0.00052$). (b) Calculation with odd helical coefficients only ($b_1,b_2,b_3,b_4,b_5,b_6$)=($0.17,0,-0.003,0, 0.0013,0$). (c) Calculation with even 'shape' coefficients only ($b_1,b_2,b_3,b_4,b_5,b_6$)=($0,-0.06,0,0.0005, 0,-0.00052$). Calculations were made in Mathematica 12.}\label{director_sk}
\end{center}
\end{figure}

It can be seen that the potential consists of two parts: the first one that depends on the chirality $U_{ij}^{helical}$ and the second depends on the envelope shape of the 3D soliton $U_{ij}^{shape}$. The set of coefficients ($b_1,b_2,b_3,b_4,b_5,b_6$) is found either from the asymptotics of the exact 3D soliton or by fitting the experimental interaction potential and/or director distribution.
Using the relation $K_{0}'(x)=-K_1(x)$ we find an expression for the force of interaction between two identical 3D skyrmions (or between two identical spiral particles of a symmetric shape with respect to $z$):

\begin{equation}
F_{ij}=F_{ij}^{helical}+F_{ij}^{shape}\label{force}
\end{equation}

\begin{equation}\label{uinthelical}
\begin{gathered}
F_{ij}^{helical}=16\pi \sigma K\sum_{n=1,3,5..}^{\infty}  K_{1}(\frac{ n\pi\rho}{L})\cdot[b_1^2\left(n\pi\sigma\right)^3-2b_1 b_3\left(n\pi\sigma\right)^5+2b_1 b_5\left(n\pi\sigma\right)^7 +\\
\\+ b_3^2\left(n\pi\sigma\right)^7-2b_3 b_5\left(n\pi\sigma\right)^9+b_5^2\left(n\pi\sigma\right)^{11} ]
\end{gathered}
\end{equation}
\begin{equation}\label{uinshape}
\begin{gathered}
F_{ij}^{shape}=16\pi \sigma K\sum_{n=2,4,6..}^{\infty} K_{1}(\frac{ n\pi\rho}{L})\cdot[b_2^2\left(n\pi\sigma\right)^5-2b_2 b_4\left(n\pi\sigma\right)^7+2b_2 b_6\left(n\pi\sigma\right)^9 +\\
\\+ b_4^2\left(n\pi\sigma\right)^9-2b_4 b_6\left(n\pi\sigma\right)^{11}+b_6^2\left(n\pi\sigma\right)^{13} ]
\end{gathered}
\end{equation}

In the work \cite{lcskyr9} the force of interaction between LC skyrmions with diameter $d=6.7\mu m$ in a cell with thickness $L=10\mu m$ and average elastic constant $K=12.1pN$ was experimentally measured. As a result of the fitting (see Fig.(\ref{pic_force})), we found that the set of coefficients can be approximately represented as ($b_1,b_2,b_3,b_4,b_5,b_6$)=($0.17,-0.06,-0.003,0.0005, 0.0013,-0.00052$). On the Fig.(\ref{director_sk}) it is shown the director field distribution with the same coefficients ($b_1,b_2,b_3,b_4,b_5,b_6$) according to formulas (\ref{nx2_sk},\ref{ny2_sk}) on the spheroidal surface with radius 0.6L with first three terms in the series (calculations were made in Mathematica 12).

It is clearly seen, that odd helical coefficients ($b_1,b_3,b_5$) define the helicity of the director field while even coefficients ($b_2,b_4,b_6$) define the envelope shape of the director field.

\section {Behavior of 3D skyrmions in a skyrmion bag}

\begin{figure}
\begin{center}
\includegraphics[width=1.0\columnwidth]{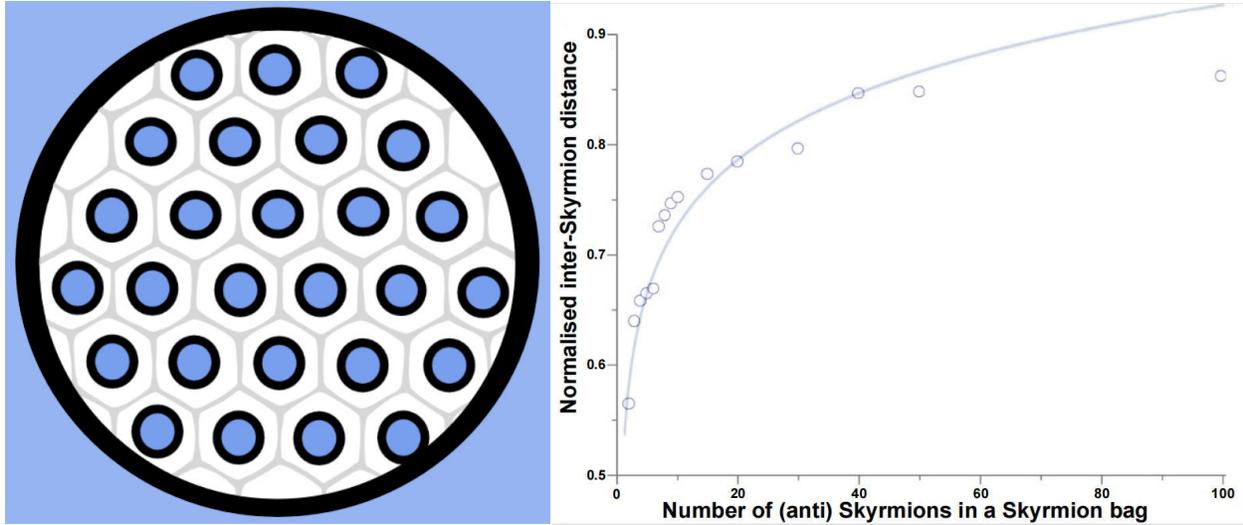}
\caption{ a) Schematic representation of a skyrmion bag with $N$ anti-skyrmions inside. The blue color schematically shows the orientation of the director downwards at an angle $\pi$, and the white color shows it upwards at an angle $0$. The black color shows the area where the director lies approximately in the horizontal plane at an angle $\pi/2$ .b) The distance between anti-skyrmions in arbitrary units depending on the number of anti-skyrmions $N$ in the bag. Experimental points were taken from \cite{lcskyr9}. Blue theoretical curve (2) is calculated according to (\ref{connection1}) with $r_1=6.2 \mu m$ , $N_1=13$ and $L=10 \mu m$. }\label{sk_bag}

\end{center}
\end{figure}

The figure (\ref{sk_bag}) schematically shows a skyrmion bag with anti-skyrmions. The blue color schematically shows the orientation of the director downwards at an angle $\pi$, and the white color shows it upwards at an angle $0$. Thus, the orientation of the director inside the small skyrmions is opposite to the orientation inside the large skyrmion bag, and therefore they can be called anti-skyrmions, although in fact they are ordinary skyrmions. The black color shows the area where the director lies approximately in the horizontal plane at an angle $\pi/2$ .
It was found in \cite{lcskyr9} that the distance between skyrmions in a bag increases with the number $N$ of skyrmions in the bag. Let's try to explain this.

It is natural to assume that the energy of the skyrmion bag itself is proportional to its circumference $E_{bag,self}=2\pi R\alpha>0$, where $R$ is the radius of the skyrmion bag (the radius of the large black circle in the figure (\ref{sk_bag} )), and $\alpha$ is the linear energy density. Inside the skyrmion bag, the skyrmions repel each other and organize a quasi-hexagonal lattice with a distance $2r$ between the centers and the unit cell area $S_{skyrmion}=2\sqrt{3}r^2$. Then $\pi R^2\approx N S_{skyrmion}$ and $r=\frac{\pi R}{2\sqrt{3N}}$. Each skyrmion interacts with 6 neighbors and then it's interaction energy is $6 U(2r)/2=3U(2r)$, where $U(2r)=a K_0(\frac{2\pi r}{L})$ is the interaction potential between two skyrmions ($a=16\pi ^3 K d b_1^2(\frac{d}{L})^3$) with $d$ being diameter of the small skyrmion (small black circles on fig. \ref{sk_bag}) and $b_1$ is the dipole multipole coefficient. We left here only the first most important dipole term from the formula (\ref{uintconfhelical1}). Then the total energy of the skyrmion bag with $N$ skyrmions inside becomes:

\begin{equation}\label{bag_energy}
E_{bag,N} (R)= 2\pi R\alpha +3 a N K_0\left({\frac{\pi^2 R}{\sqrt{3N}L}}\right)
\end{equation}

The equilibrium condition $\frac{\partial}{\partial R}E_{bag,N}=0$ leads to the equation:

\begin{equation}\label{bag_eq}
\frac{2\alpha L}{\sqrt{3N}a}=K_1\left({\frac{\pi^2 R}{\sqrt{3N}L}}\right)
\end{equation}

Using the asymptotics $K_1(x)\approx e^{-x}\sqrt{\frac{\pi}{2x}}$ for big $x$ and introducing the dimensionless variable $x=\frac{2\pi r}{ L}$ we can get such a relationship between different pairs $(N_1,x_1)$ and $(N_2,x_2)$:

\begin{equation}\label{connection}
x_2-x_1+\frac{1}{2}ln\left(\frac{x_2}{x_1}\right)=\frac{1}{2}ln\left(\frac{N_2}{N_1}\right)
\end{equation}

Suppose we know from the experiment the distance $x_1$ for one skyrmion bag with $N_1$ skyrmions inside. Then it is required to define $x_2$ for any other $N_2$. First, consider the case when $|ln(\frac{N_2}{N_1})|<1$. Then we introduce $\epsilon=\frac{x_2}{x_1}-1$ and expand the equation (\ref{connection}) to $\epsilon^2$. As a result, we get the following solution for the size $r_2$ of one elementary cell of the bag with $N_2$ skyrmions:

\begin{equation}\label{connection1}
r_2=r_1+r_1\cdot\left(1+\frac{4\pi r_1}{L}-\sqrt{\left(1+\frac{4\pi r_1}{L}\right)^2-2ln\left(\frac{N_2}{N_1}\right)} \right)
\end{equation}

and the distance between two nearest skyrmions will be $2r_2$. It is known from the experiment \cite{lcskyr9} that $r_1=6.2 \mu m$ for the bag $S_{13}$ with $N_1=13$ and $L=10 \mu m$. Then we get the normalized curve $r_2(N_2)$ in the figure (\ref{sk_bag},b). It can be seen that it describes quite well the experimental points for different numbers of (anti) skyrmions.

One can also get the theoretical asymptotics for $N_2\gg N_1$, then:

\begin{equation}\label{connection2}
r_2=r_1+\frac{L}{4\pi}ln\left(\frac{N_2}{N_1}\right)
\end{equation}

\section{Conclusion}

In this paper, we tried to bridge the gap between 3D skyrmions and colloidal particles in liquid crystals. We looked at a 3D skyrmion in a cholesteric homeotropic cell as a colloidal particle and considered the interaction between 3D skyrmions as an elastic interaction between colloidal particles. This can be done exactly outside the core of the skyrmion, beyond the region with strong director deformations. Then the director distribution can be accurately determined analytically and the elastic interaction potential between 3D skyrmion particles can be calculated exactly. We considered skyrmions with axial symmetry in a homeotropic cell of thickness $L$ with strong boundary conditions, when the director looks vertically at the surfaces of the bounding planes. It turns out that the chirality of the liquid crystal manifests itself in the birth of chiral multipole coefficients $a_l^{\gamma}$ , which are absent in ordinary colloidal particles in the nematic. Usually, colloidal particles in nematics have only zero-helicity multipole coefficients $a_l^{0}$. And then we show that the 3D skyrmion in the region of weak deformations can be described as a set of multipoles: zero $a_l^{0}$ and chiral $a_l^{\gamma}$. But we do not need to know all $a_{l}^{\gamma}$, as it is sufficient to know only two coefficients $(a_l^x,a_l^y)$ for each $l$ to determine the director field: $(a_l^x,a_l^y)=(\sum_\gamma a_{l}^\gamma cos\gamma,-\sum_\gamma a_{l}^\gamma sin\gamma)$.

We have considered in more detail the case of 3D skyrmion particles which have symmetrical shape with respect to the median plane $z=L/2$ and consider only the helicity $\gamma=\pi/2$. We also limited ourselves to multipole moments up to the 6-th order inclusively. Then we obtained the interaction potential between 3D skyrmions and found the multipole coefficients to be in agreement with the experiment.

Next, we applied our results to 3D skyrmions in a skyrmion bag. As the number of skyrmions $N$ in the bag increases, the distance between them gets larger as well. We have found an approximate analytical formula for the distance between skyrmions depending on the number of $N$ skyrmions in the bag, which also agrees with experiment.

Along the way, we determined the director field around a chiral axially symmetric colloidal particle in the unlimited nematic liquid crystal and in the confined homeotropic cell, taking into account higher order elastic terms with orders $l=1\div 6$. In general case it can have as well a double set of multipoles: zero $a_l^{0}$ and chiral $a_l^{\pi/2}$ for $l=1\div 6$. We also have found the elastic interaction potential between such particles.

\end{document}